\documentclass[10pt,letterpaper]{article}
\usepackage{opex3}

\usepackage{cite}
\usepackage{amsmath,amssymb,graphicx, color, verbatim}

\DeclareGraphicsExtensions{.pdf, .png, .jpg}

\begin{document}

\title{Compact, CO$_2$-stabilized tuneable laser at 2.05 microns}

\author{Philip G. Westergaard$^{1,*}$, Jan W. Thomsen$^2$, Martin R. Henriksen$^2$, Mattia Michieletto$^3$, Marco Triches$^1$, Jens K. Lyngs\o$^3$, and Jan Hald$^1$} 
\address{
$^1$Danish Fundamental Metrology, Matematiktorvet 307, DK-2800 Kgs. Lyngby, Denmark\\
$^2$Niels Bohr Institute, Blegdamsvej 17, DK-2100 Kbh. \O, Denmark\\
$^3$NKT Photonics A/S, Blokken 84, DK-3460 Birker\o d, Denmark }

\email{$^*$Corresponding author: pgw@dfm.dk}

\begin{abstract}
We demonstrate a compact fibre-based laser system at 2.05 microns stabilized to a CO$_2$ transition using frequency modulation spectroscopy of a gas-filled hollow-core fibre. The laser exhibits an absolute frequency accuracy of 5~MHz, a frequency stability noise floor of better than 7~kHz or $5 \times 10^{-11}$ and is tunable within $\pm 200$ MHz from the molecular resonance frequency while retaining roughly this stability and accuracy. 
\end{abstract}


\section{Introduction}
The development of hollow-core photonic crystal fibre (HC-PCF) technology over the past decade has opened up a vast array of possibilities for new applications. When filled with gas, the HC-PCF is a good candidate for frequency stabilization and sensing in demanding environments such as field operations and space applications, since this provides a compact, low-weight container for molecular gas while retaining large interaction length between laser light coupled trough the fibre and the molecules inside the hollow core. Specifically for satellite-based LIDAR (LIght Detection And Ranging) measurements, a lightweight laser source with high stability and accuracy is desirable, since the accuracy of the LIDAR measurement depends directly on these parameters.

Current state-of-the-art laser frequency stabilisation using gas-filled hollow-core fibres have demonstrated a relative long-term instability of about $4\times 10^{-12}$ and relative short-term instability of about $2 \times 10^{-12}$ for one second averaging time \cite{Marco1, Lurie, Corwin}. These results were obtained with acetylene and iodine using saturated absorption spectroscopy, which gives linewidths of typically 20 MHz. The main limitations to the stability are decoherence due to collisions between molecules and the inner wall of the HC-PCF as well as excitation of higher order optical modes in the HC-PCF. Stabilization to Doppler broadened absorption lines has also been demonstrated in the telecom wavelength range using acetylene \cite{Benabid} as well as in the 2 micron range using CO$_2$ filled HC-PCF \cite{Meras}, but without the possibility of tuning the offset frequency.

Here, we report on a compact fibre-based laser system at 2.05 microns stabilised to the CO$_2$ P(30) transition at $2050.967$~nm using frequency modulation spectroscopy \cite{Bjorklund} in a gas-filled hollow-core fibre. The laser system was developed as a part of a project for the European Space Agency (ESA) aiming towards satellite-based LIDAR measurements. Several requirements on size, accuracy and stability of the system  were therefore imposed; all of which our system meets. The compact fiber-based optical setup fits inside a box with dimensions $(25 \times 25 \times 5)$~cm$^3$. The laser exhibits an absolute frequency accuracy of better than 5 MHz or $3 \times 10^{-8}$, frequency stability of better than $10^5 \textrm{ Hz}/\sqrt{\tau}$ up to around 1000~s and is tunable within $\pm 200$ MHz from the molecular resonance frequency while retaining roughly this stability and accuracy. 

\section{Experimental Setup}
\begin{figure}[!h]
\centering\includegraphics[width=14cm]{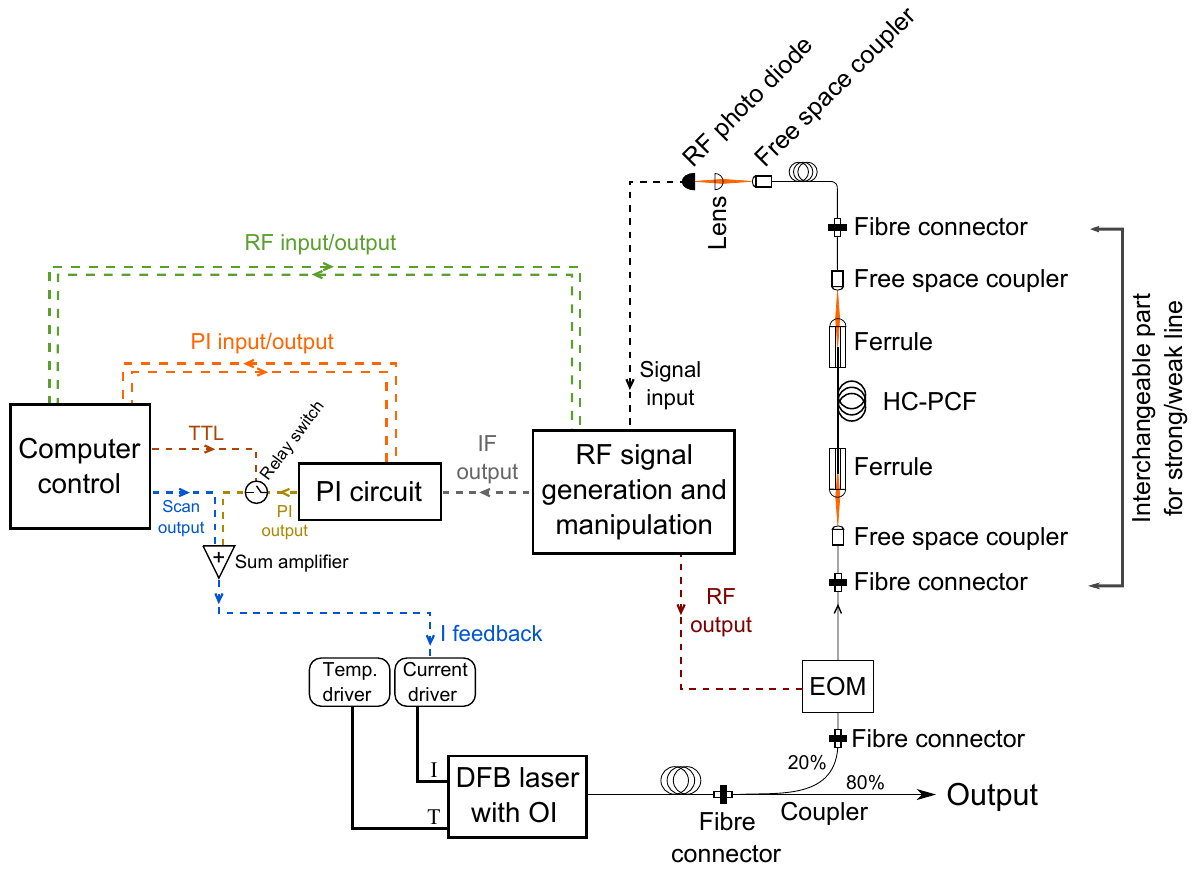}
\caption{An overview of the experimental setup. OI: Optical Isolator, HC-PCF: Hollow-Core Photonic Crystal Fiber. The dashed lines indicate electronic DC or RF signal paths.}
\label{Fig:Setup}
\end{figure}
The optical part of the setup consists of a fiber-coupled DFB laser (EP2051 series, Eblana Photonics), an electro-optic modulator (EOM), two HC-PCFs of different length for probing different molecular transitions, two ferrules per HC-PCF for gas filling and coupling of light in and out of the HC-PCF, and a photo detector. 

An overview of the experimental setup is shown in Fig.~\ref{Fig:Setup}.
The DFB laser comes in a polarization-maintaining fiber-coupled butterfly package. Furthermore, a thermo-electric element for temperature control at the mK level is integrated in the laser package. The light is split after the laser using a fiber splitter, where 80 \% is used as output light (4.3~mW) and 20 \% is used to generate an error signal for locking. The part of the light used for locking is first phase modulated at frequencies between 300 MHz and 3 GHz using a fiber-coupled EOM (MPZ2000 series, Photline Technologies).

After modulation, the light is sent through the HC-PCF. The coupling between the PM/SM fibres used for the laser and the HC-PCF utilizes a ferrule method developed in-house at DFM \cite{Marco}. The HC-PCF is inserted into one end of a glass ferrule and an aspheric lens is attached to the other end for coupling. To ensure optimal, alignment-free coupling the PM/SM fibre is connected to a free space coupler, which is bonded together with the HC-PCF ferrule. The alignment and coupling are optimised using a 6-axis translational/rotational stage before gluing. A glass tube attached to the side of the ferrule is used to first evacuate and then fill the HC-PCF to the desired pressure. After filling the system is sealed by melting the side tube shut with a blow torch.
After travelling through the HC-PCF, the light is coupled into free space again and detected on the high bandwidth (12~GHz) photo diode. Around $50\;\mu$W of the light reaches the detector when the laser is tuned off resonance. 

\begin{figure}[!h]
\centering\includegraphics[width=8cm, angle=0]{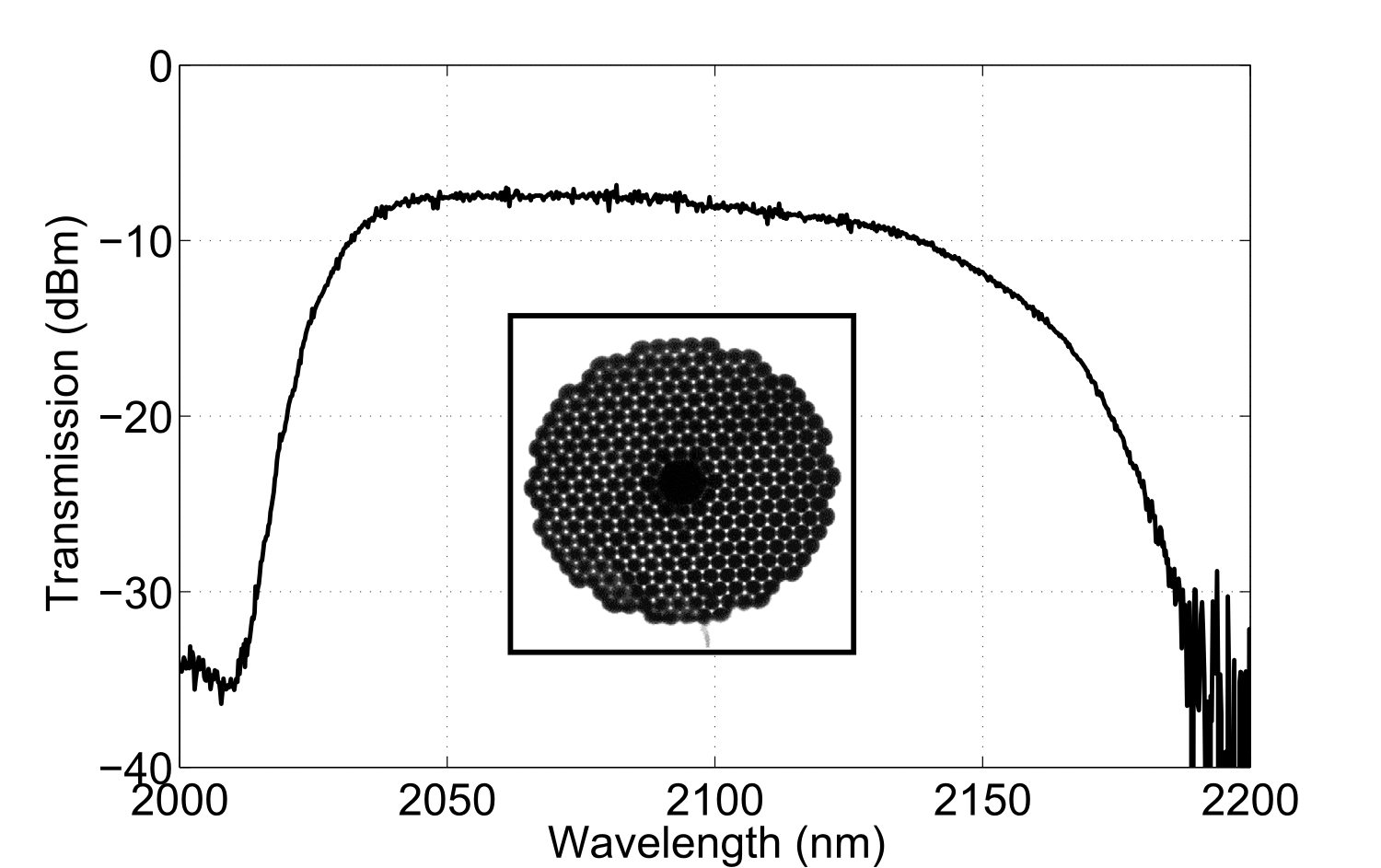}
\caption{Characteristics of the fabricated HC-PCF. Transmission measurement through 1~m fiber. The inset shows a microscope image of the end face of the HC-PCF.}
\label{Fig:Fiber}
\end{figure} 
The HC-PCF employed in our setup is specifically designed to increase the frequency stability of the laser lock. Mode interference in the HC-PCF has a negative influence on the frequency stability, since it causes a varying dependence of the transmission on laser offset frequency, which distorts the error signal of the lock unpredictably. Therefore, we designed and fabricated a fiber for guidance around 2050~nm with suppression of higher order core modes. The fiber is a 7cell HC-PCF with reduced core wall thickness to enable single mode propagation at the short wavelength region of the bandgap, similar to the one described in \cite{Lyngsoe}. It has a cladding air filling fraction of approximately 90 percent, a core diameter of $15\;\mu$m and a cladding pitch of $5\;\mu$m. Figure~\ref{Fig:Fiber} shows the 150~nm wide transmission window centered at 2100~nm. 
\begin{figure}[!h]
\centering\includegraphics[width=8cm, angle=0]{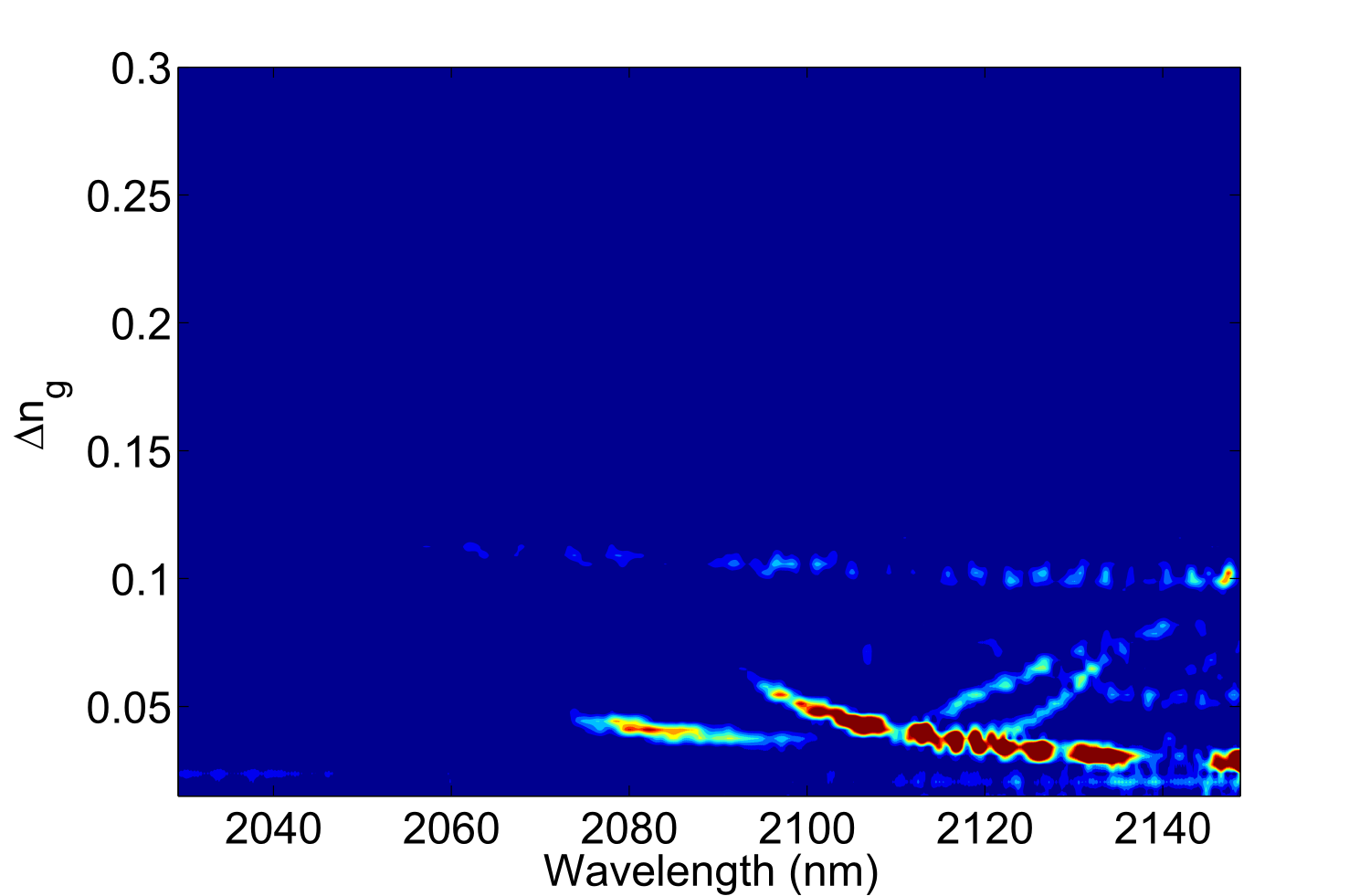}
\caption{Higher order mode characterization of the HC-PCF using Windowed Fourier Transform method. The figure shows the mode content of the differential group indices with red being the densest and blue the thinnest. Dense mode content for non-zero differential group index indicates the presence of higher order modes in the fiber.}
\label{Fig:Fiber_HOM}
\end{figure} 
Characterization of the higher order mode content can be achieved using white light interferometric measurements as described in references \cite{Gerosa, Marco1}. Using the same setup used in the characterization of the HC-PCFs presented in \cite{Marco1}, a windowed Fourier transform analysis can be used to temporally separate the different modes and determine their effective group index difference with respect to the fundamental mode. The high order modes travel with a different group velocity along the HC-PCF with respect to the fundamental mode. This results in a relative delay of the wave fronts between the modes which causes an interference occurring between the majority of the light coupled into the fundamental mode and the other modes. A windowed Fourier transform analysis allows one to separate the different modes in terms of the group velocity different with respect to the fundamental one.
To quantify this, light from a white light source was launched with high NA into a 50~cm section of the HC-PCF. At the output end of the HC-PCF light was collected with a single mode fiber and the resulting interferogram was recorded using a Yokogawa AQ6375 spectrum analyzer. Figure~\ref{Fig:Fiber_HOM} shows the differential group indices calculated from the windowed Fourier transform of the interferogram. Here, it can be seen that the current fiber has a higher order mode cut-off at around 2070~nm, below which the fiber only supports a single spatial core mode. The relatively flat response at a differential group index of approximately 0.1 is an artefact of the measurement equipment similar to the one observed in \cite{Marco1}.

\begin{figure}[!h]
\centering\includegraphics[width=11cm]{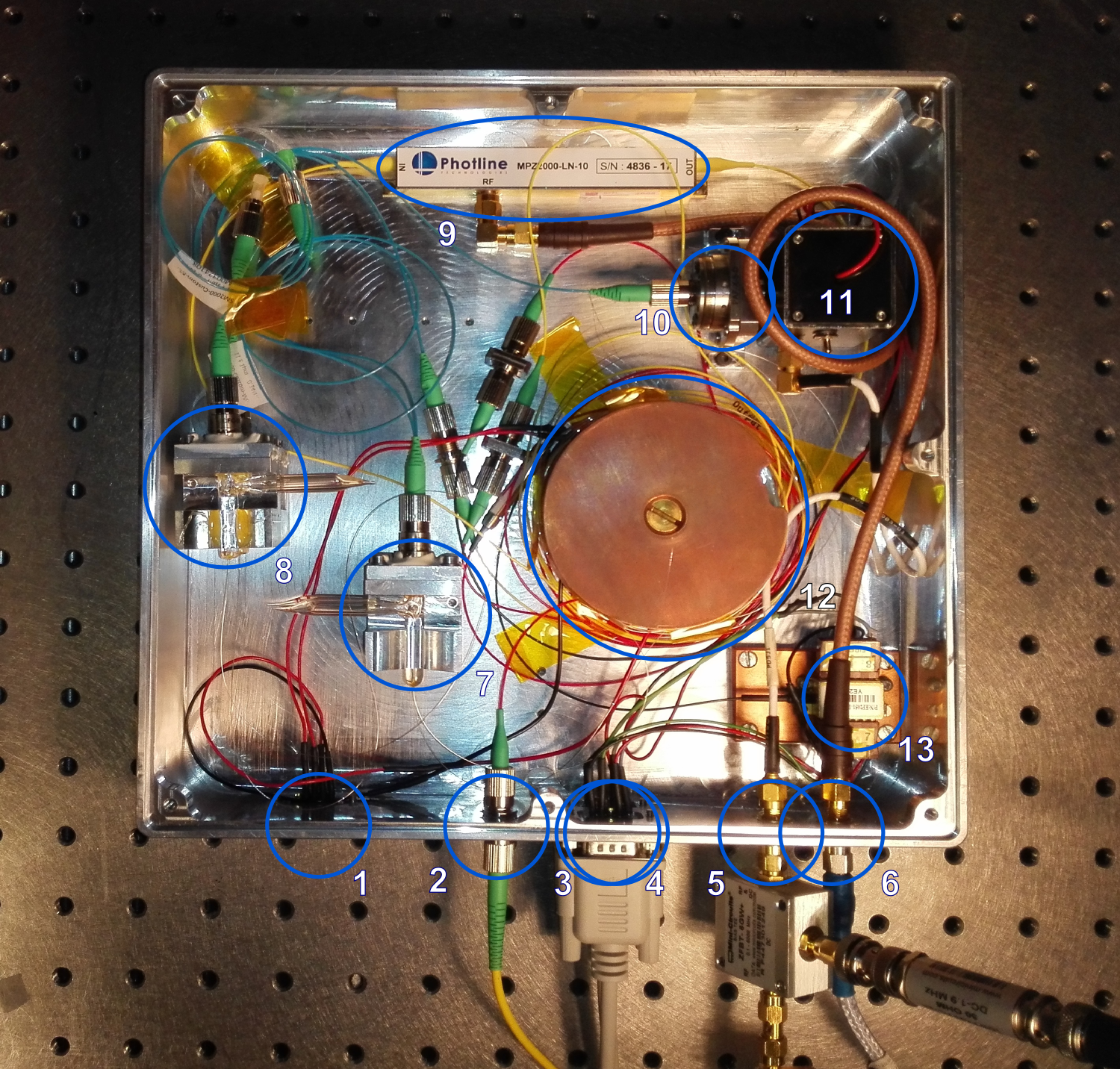}
\caption{An overview of the optical setup inside a box with outer dimensions $(25 \times 25 \times 5)$~cm$^3$. The numbered items are described in Table \ref{Tab:OpticalBox}.}
\label{Fig:ESA_box}
\end{figure}

\begin{table}[!h]
    \begin{center}
		\caption{\small Explanations of the numbered items in Fig. \ref{Fig:ESA_box}.}
    \begin{tabular}[b]{| c | l |}
        \hline
				\textbf{Designation} & \textbf{Description} \\
        \hline
				1  & Input/output for the temperature control of the fibre spools. \\
				$\;$ & This includes the Peltier element and the thermistor hidden beneath the fibre spools. \\
				2 & Optical output (80 \% arm from fibre coupler). \\
				3 & (top) Electrical connections for the current and temperature control of the DFB laser. \\
				4 & (bottom) 3V supply for the biased photo diode. \\
				5 & RF and DC output from the photo diode. \\
				$\;$ & The RF and DC part is split with a biastee outside the box. \\
				6 & RF input for driving the EOM. \\
				7 & Input and output ferrules for the short HC-PCF (mounted on top of each other). \\
				8 & Input and output ferrules for the long HC-PCF (mounted on top of each other). \\
				9 & Electro-Optics Modulator. MPZ2000 series, Photline Technologies. \\
				10 & Free space coupler with additional lens for coupling into photo detector. \\
				11 & Biased photo diode. Newport 818-BB-51. \\
				12 & Fibre spools that hold the short (bottom) and long (top) HC-PCFs. \\
				13 & DFB laser on copper mount.\\				
				\hline
    \end{tabular}
    \label{Tab:OpticalBox}
    \end{center}
\end{table}
The complete optical part of the setup fits in a compact box with dimensions $(25 \times 25 \times 5)$~cm$^3$. The assembled optical box is presented in Fig.~\ref{Fig:ESA_box}.
The numbered components in the figure are described in Table \ref{Tab:OpticalBox}.

For increased versatility, we have produced two different HC-PCFs of different length, both coupled, filled and sealed using the ferrule technique. Fiber 1 has a length of $L_1 = 10$~m and was filled with $^{12}$CO$_2$ to a pressure of $P_1 = 35$~hPa and the second fiber has $L_2 = 100$~m and filled to $P_2= 55$~hPa with $^{13}$CO$_2$. Each fiber is coiled on a copper spool of 8~cm radius and the two spools are placed on top of each other (see item 12 in Fig. \ref{Fig:ESA_box}). The pressure in the short fiber was selected to optimize the error signal slope for the strong CO$_2$ line at 2050.967~nm. Similarly, the pressure in the long fiber was optimised for a different transition at 2050.7222~nm which has a line strength roughly a factor 100 weaker than the strong line. Both fibers can be used in the optical setup by a simple change of input/output fiber connectors in the optics box. In this work, we will concentrate on the results obtained from the short fiber with the strong transition at 2050.967~nm.

\begin{figure}[!h]
\centering\includegraphics[width=13cm]{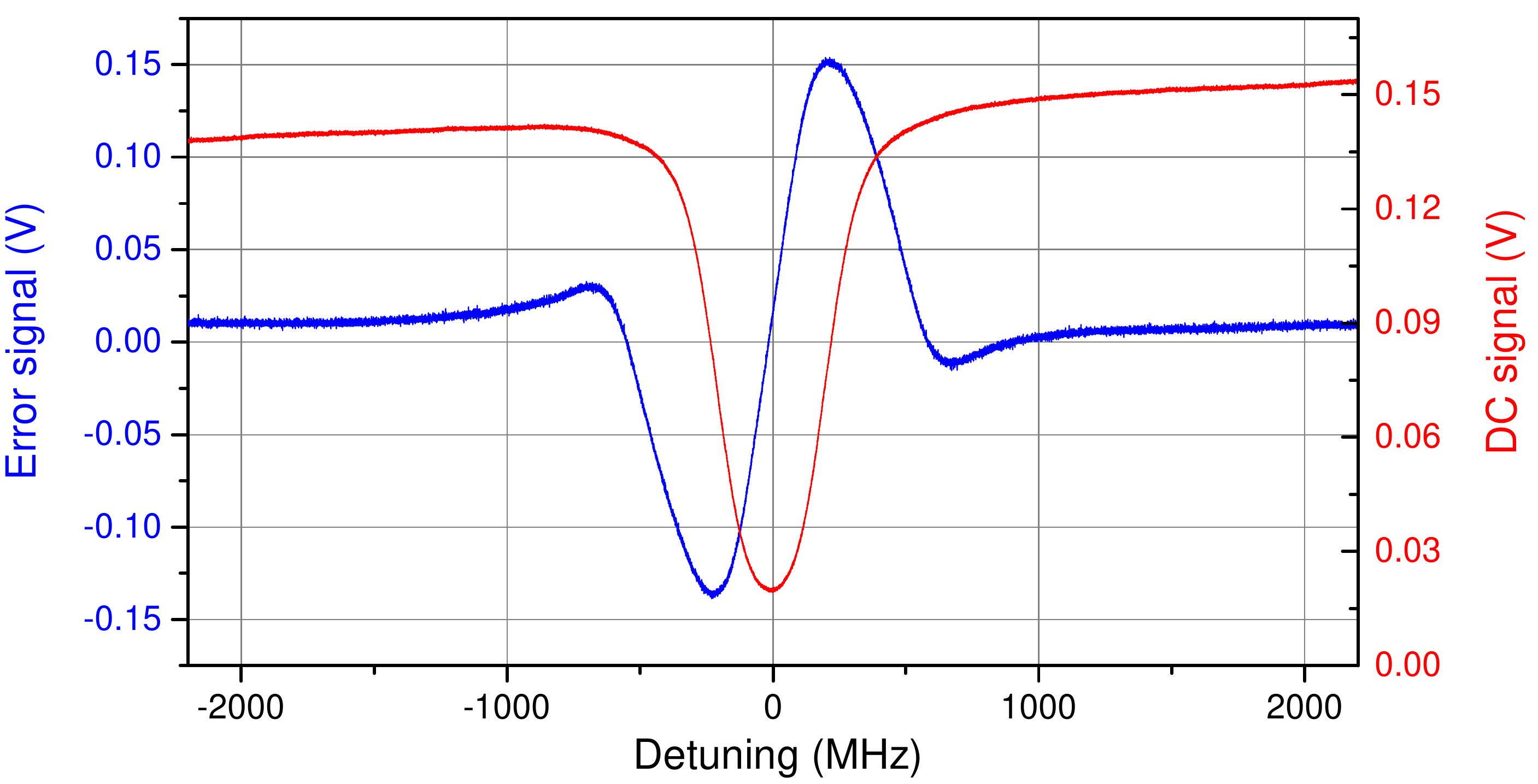}
\caption{The absorption signal (red curve) derived from the photo detector DC level, and the corresponding FM spectroscopy error signal (blue curve). The sweep time was 2 seconds without any averaging.}
\label{Fig:ErrSign}
\end{figure}

The drive current and temperature regulation for the DFB laser is supplied to the optical assembly from an external box containing all drive and control electronics (not shown in Fig. \ref{Fig:ESA_box}). Similarly, the RF signal driving the EOM is sent from the electronics box to the RF input of the optical assembly. The signal from the light detected on the RF photo diode is split into DC and RF parts by a bias-tee outside the box, and the RF signal is amplified before being mixed with the LO at the modulation frequency. This produces an error signal used for locking. The DC signal shows the molecular absorption and is used for frequency calibration and monitoring of the lock status. The DC and error signals are shown in Fig. \ref{Fig:ErrSign}.

The design also features the possibility to temperature stabilize the spool holding the HC-PCF (not shown in Fig. \ref{Fig:Setup}). This is done using a Peltier element inserted between the spool and the bottom of the optics box. In the stable laboratory environment, however, it was found unnecessary to stabilize the temperature. In a more noisy setting, stabilizing the temperature might prove beneficial to reduce fluctuating transmission variations of the HC-PCF. 

The signal processing and locking is controlled by a computer program through a National Instruments DAQ board. The program can select the modulation frequency from one of two VCOs in the ranges [250, 400] MHz and [1800, 3200] MHz. The program features the possibility to lock off resonance at a user-determined offset frequency in the ranges [-200, 200] MHz and [$\pm 1900, \pm 2800$] MHz. For the low frequency offset an electronic offset is added to the error signal obtained with low frequency modulation ($\sim 320$~MHz, as shown in Fig. \ref{Fig:ErrSign}) before the PI circuit. This way, instead of locking at the frequency corresponding to the zero crossing of the error signal, the PI circuit will lock the laser to value of the error signal corresponding to the electronic offset. The relation between the electronic offset and the resulting offset frequency is obtained by a calibration of the frequency axis in the control program. This calibration sequence takes place automatically in the program and occurs at user-determined time intervals. 

The first step in the calibration sequence is determining the zero detuning. The low frequency VCO is set to the frequency where the last calibration found the optimum (i.e., largest) slope of the error signal. The laser current is then scanned by sending a voltage ramp to the modulation input of the current driver from the DAQ board. From here, the slope is optimised again by changing the VCO frequency slightly. The slope is obtained by a linear fit around the last zero crossing value.
Once the largest slope is determined, the scan voltage corresponding to zero frequency offset is determined by the scan voltage where the error signal crosses zero (corrected by a possible offset determined from residual amplitude modulation (RAM)).
After the zero offset scan voltage has been determined, the high frequency VCO is used to obtain the scan voltage values for higher offset frequencies. The VCO is set to (2000, 2250, 2500, 2750, 3000) MHz and for each frequency, the corresponding scan voltage is found from the apogee of a parabolic fit to a zoom on the DC signal's positive and negative sidebands from the EOM modulation. 
After this procedure, eleven pairs of frequency offsets and scan voltage are found, and a third order polynomium is fitted to the points to relate the scan voltage to a frequency offset. 

\section{Results}
\begin{figure}[!ht]
\centering\includegraphics[width=12cm]{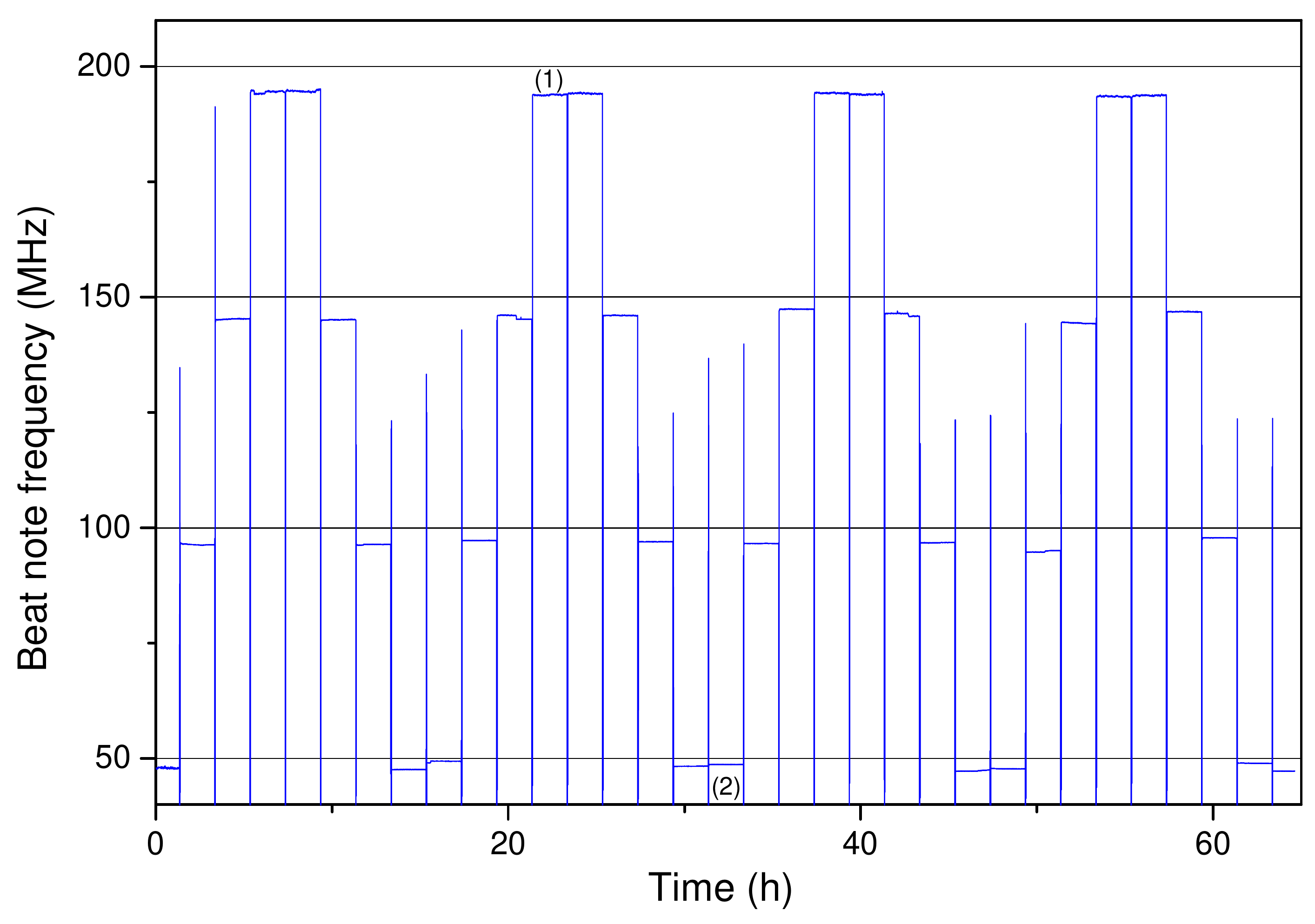}
\caption{The time series of the beat note frequency between the system investigated here and a reference laser locked at resonance to a CO$_2$-filled Brewster angled glass cell. The offset frequency is changed every two hours with a re-calibration taking place before changing the offset frequency. The initial offset is +50 MHz and is increased in steps of +50 MHz after each re-calibration. After reaching +200 MHz the offset is set to -200 MHz. Note that negative offset frequencies are also recorded as a positive beat note frequency. The numbers (1)
and (2) refer to selected periods for which the Allan deviation is plotted in Fig.~\ref{Fig:AllanDev}.}
\label{Fig:TimeSeries}
\end{figure}

Once the calibration procedure is completed the laser can be locked at an offset frequency selected by the user. The control program will then add a suitable electronic offset to the PI circuit found from the calibration to lock the laser to the corresponding frequency offset from resonance.
The time series of a 65 hour measurement run with different offset frequencies is shown in Fig.~\ref{Fig:TimeSeries}.
The data was obtained from a frequency counter measuring the beat note between a reference laser and the system described here.

\begin{figure}[h]
\centering\includegraphics[width=12cm]{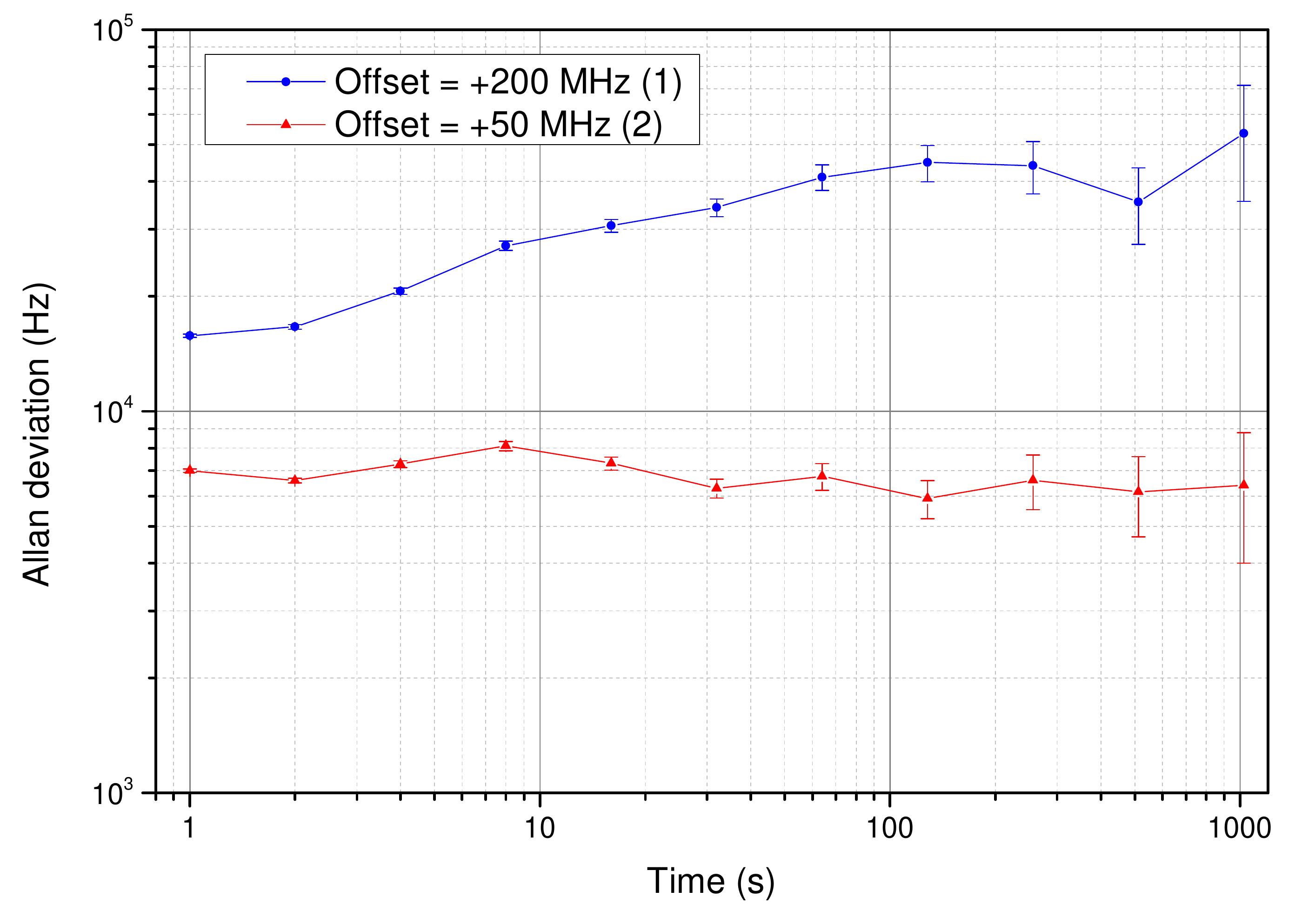}
\caption{The Allan deviation obtained at two different offset frequencies (marked with (1) and (2) in Fig.~\ref{Fig:TimeSeries}).}
\label{Fig:AllanDev}
\end{figure}

Since the effective error signal slope decreases away from resonance, large
offsets typically display a higher instability than small offsets. Figure~\ref{Fig:AllanDev} shows the Allan deviation
of two representative measurements of +50 MHz and +200 MHz offset frequencies marked by (1) and (2) in Fig.~\ref{Fig:TimeSeries}.
On the times scale considered here (1 - 1000 s), both offset frequencies are mostly dominated by flicker noise floor, typically originating from electronic components - the +50 MHz offset at a level of around $6.5$ kHz, or $4.4 \times 10^{-11}$ in relative units. The increased level of flicker noise in the +200 MHz offset for small time scale (around 1 s) corresponds roughly to the decrease in the error signal slope at this offset (and therefore also a decrease in the signal-to-noise ratio of the lock). At later times the +200 MHz shows beginning signs of frequency random walk (the Allan deviation increasing with $\tau^{1/2}$) and frequency drift (increasing with $\tau$), likely due to systematic effects such as baseline variations. 

\begin{figure}[h]
\centering\includegraphics[width=12cm]{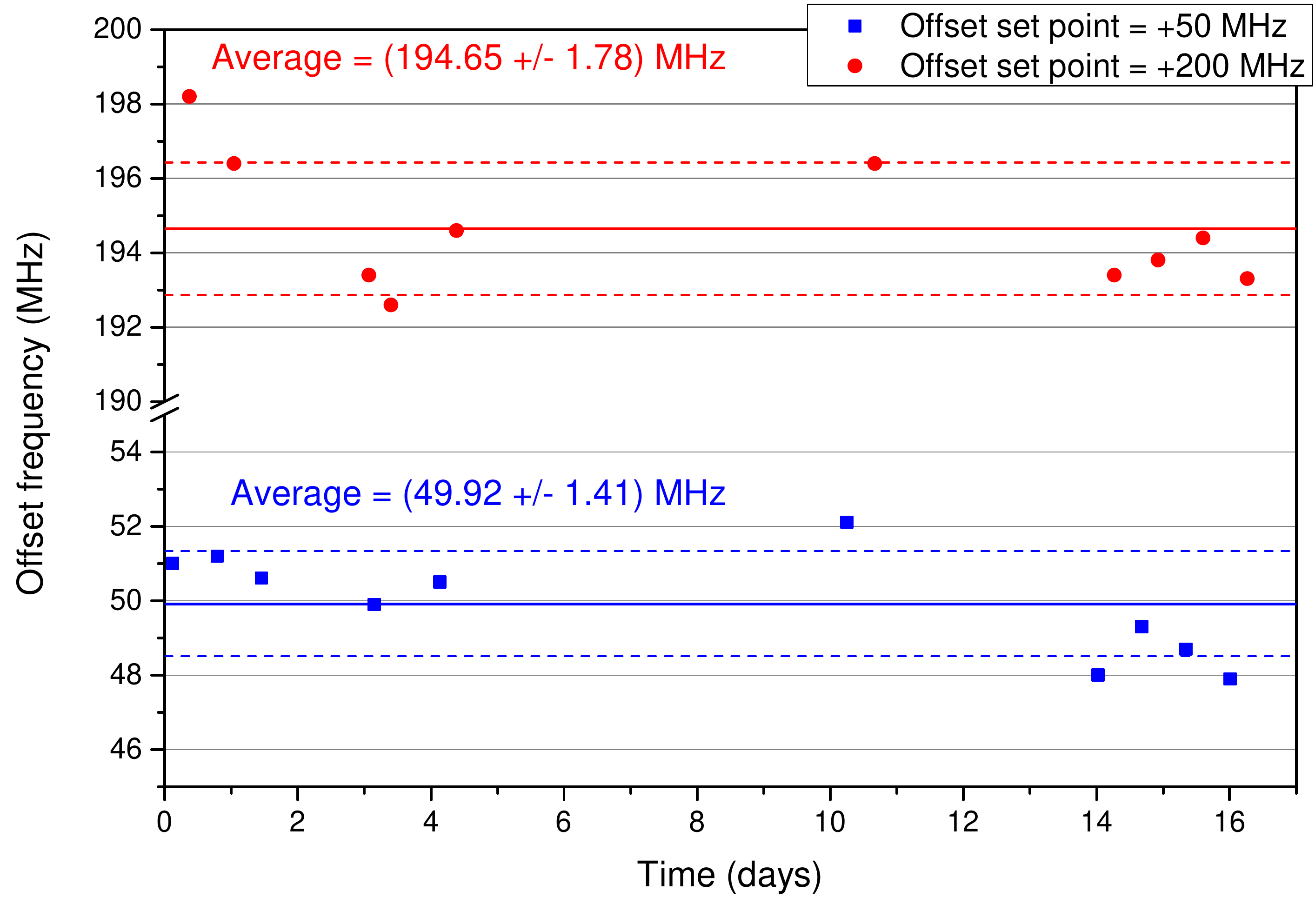}
\caption{Measurements over more than two weeks of the actual offset frequency value for +50 MHz and +200 MHz set points. The solid coloured lines show the average value and the colored dashed lines show +/- one standard deviation from the average value.}
\label{Fig:Reprod}
\end{figure}

Figure~\ref{Fig:Reprod} shows a compilation of the average value of actual offset frequencies for +50 MHz and
+200 MHz set points spanning a total of 16 days. The overall average value with standard deviation is noted for both +50
MHz and +200 MHz in Fig.~\ref{Fig:Reprod}.
While the low frequency offset value is quite close to the set point, the high frequency value is not. By analyzing different offsets we observed that the difference from the set point scales linearly with offset frequency and is due to a 4~\% underestimation of the scaling factor from the scan voltage to frequency offset calibration, most likely due to non-linearity in the laser scan during calibration. This can be compensated for by correcting the offset in the control software. If such a compensation is carried out, the actual offset frequency will be within $\pm 2$~MHz of the selected value. The consistency and accuracy of the offset lock is a result of the inherent stability of the error signal amplitude due to the compact and robust optical system. Combining this with the automatic calibration procedure ensures that the added electronic offset always corresponds to the desired frequency.

\newpage
\section{Conclusion}
We have demonstrated a compact, fiber-based setup for frequency locking of a DFB laser to a CO$_2$ transition at 2050.967~nm. The frequency of the laser can be locked up to $\pm 200$~MHz away from resonance without the need of a second laser, thus reducing the complexity of the setup - something especially useful in space applications. An automated calibration sequence ensures the long term accuracy of the offset frequency. Close to resonance the accuracy of the locked laser frequency is better than 2~MHz with a stability of 6.5~kHz between 1 and 100 seconds. The current limitation to the short term stability is the signal-to-noise ratio in the error signal used for locking. This could be improved with more optical power, and a better detection/amplification system. The long term accuracy is mostly dominated by baseline variations in transmission of the fiber as well as residual amplitude modulation from RF pickup in the amplification chain.

When baseline variations in the fiber transmission are small, the hollow-core fiber is a well-suited gas cell for locking to weak Doppler broadened molecular lines, since it is possible to achieve a very long path length (up to at least 100 m) while not being limited by wall collisions and/or transit time broadening. The gas filled into the fiber is not limited to CO$_2$, it can be filled with a large range of molecules relevant for, e.g., atmospheric measurements.

\section*{Acknowledgments}
This work was supported by the European Space Agency (ESA) under ESA Contract No. 4000107880/13/NL/PA, the Marie Curie Initial Training Network QTea - Quantum
Technology Sensors and Applications, financed by the FP7 program of the European Commission (contract-N MCITN-317485), and supported by funds from the The Danish Agency for Science, Technology and Innovation.

\end{document}